# Long-Range Structural Order in a Hidden Phase of Ruddlesden–Popper Bilayer Nickelate $La_3Ni_2O_7$

Haozhe Wang[1], Long Chen[2], Aya Rutherford[2], Haidong Zhou[2], Weiwei Xie[1]*

1. Department of Chemistry, Michigan State University, East Lansing, MI, 48824, USA
2. Department of Physics and Astronomy, University of Tennessee, Knoxville, TN, 37996, USA

* Email: xieweiwe@msu.edu

## Abstract

The recent discovery of superconductivity in Ruddlesden–Popper bilayer nickelate, specifically $La_3Ni_2O_7$, has generated significant interest in the exploration of high-temperature superconductivity within this material family. In this study, we present the crystallographic and electrical resistivity properties of two distinct Ruddlesden-Popper nickelates: the bilayer $La_3Ni_2O_7$ (referred to as 2222-phase) and a previously uncharacterized phase, $La_3Ni_2O_7$ (1313-phase). The 2222-phase is characterized by a pseudo *F*-centered orthorhombic lattice, featuring bilayer perovskite [$LaNiO_3$] layers interspaced by rock salt [$LaO$] layers, forming a repeated …2222… sequence. Intriguingly, the 1313-phase, which displays semiconducting properties, crystallizes in the *Cmmm* space group and exhibits a pronounced predilection for a *C*-centered orthorhombic lattice. Within this structure, the perovskite [$LaNiO_3$] layers exhibit a distinctive long-range ordered arrangement, alternating between single- and trilayer configurations, resulting in a …1313… sequence. This report contributes to novel insights into the crystallography and the structure-property relationship of Ruddlesden–Popper nickelates, paving the way for further investigations into their unique physical properties.

## Introduction

Nickelates have emerged as promising candidates for the exploration of additional high-temperature superconductors and for understanding the origins of high $T_C$ superconductivity in cuprate superconductors reported in the 1980s.[1] This interest stems from their analogous crystal and electronic structures.[2] Two major families are Ruddlesden–Popper phases and reduced square-planar phases. Ruddlesden–Popper phases feature perovskite layers with octahedral coordination [$NiO_6$] separated by rock salt layers, represented by the general formula $R_{n+1}Ni_nO_{3n+1}$ ($n$ = 1, 2, 3, …, ∞).[3,4] On the other hand, reduced square-planar phases exhibit square net layers with square-planar coordination [$NiO_4$], where rock salt layers transform to fluorite blocking slabs, summarized by the general formula $R_{n+1}Ni_nO_{2n+2}$ ($n$ = 1, 2, 3, …, ∞).[5-9]

Previously, superconductivity has been reported in epitaxial thin films of reduced square-planar phases, including infinite-layer $Nd_{0.8}Sr_{0.2}NiO_2$[10], $Pr_{0.8}Sr_{0.2}NiO_2$[11,12], $La_{1-x}Sr_xNiO_2$[13], $La_{1-x}Ca_xNiO_2$[14], and quintuple-layer $Nd_6Ni_5O_{12}$[15,16]. These systems feature ultra-low valence $Ni^{1+}$, isostructural to $Cu^{2+}$, and the induction of hole-doping suggests that superconductivity stabilizes with a similar formal $3d$ electron count. However, the synthesis of these epitaxial thin films through topotactic reduction with metal hydrides introduces the possibility of hydrogen doping, casting ambiguity on the origin of superconductivity.

Recently, a groundbreaking discovery reported superconductivity in the Ruddlesden–Popper bilayer nickelate $La_3Ni_2O_{7-\delta}$, achieving a $T_C$ up to 80 K in the pressure range of 14 GPa to 43.5 GPa,[17] sparking significant interest in Ruddlesden–Popper phases ($n$ = 2 and $n$ = 3). In the ambient pressure structure of Ruddlesden–Popper bilayer $La_3Ni_2O_{7-\delta}$, the $NiO_6$ octahedra display rotation/tilt alignment along the $c$-axis, deviating from the regular square net observed in high $T_C$ cuprates. It has been proposed that a structural transition from ambient pressure *Amam* to the high-pressure *Fmmm* phase occurs at approximately 14 GPa, coinciding with the onset of superconductivity. However, the high pressure phase remains unclear due to the resolution limitations of the reported powder X-ray diffraction (XRD) data. Additionally, the potential presence of oxygen vacancies introduces sample-dependent variability in multiple physical property measurements.[17-21] The structural ambiguity and the crucial role of the structure-property relationship in comprehending the origins of high $T_C$ superconductivity necessitate more focused attention and efforts in the study of the crystal structure in this system.

Herein, we present the crystal structure and electrical resistivity of two Ruddlesden–Popper nickelates: bilayer $La_3Ni_2O_7$-2222, and a hidden phase, $La_3Ni_2O_7$-1313. Single crystal XRD is employed to determine their crystal structures. Our measurements confirm the pseudo $F$-centered orthorhombic lattice and its Ruddlesden–Popper bilayer stacking in $La_3Ni_2O_7$-2222. Remarkably, $La_3Ni_2O_7$-1313 adopts a strongly preferred $C$-centered orthorhombic lattice with the space group $Cmmm$. In this structure, perovskite $[LaNiO_3]$ layers exhibit a systematic long-range order, alternating between single- and trilayer configurations (…1313…). This report introduces new possibilities for exploring the crystal structure and the structure-property relationship in Ruddlesden–Popper nickelate superconductors.

## Experimental

**Materials growth**: Crystals of $La_3Ni_2O_7$-2222 and $La_3Ni_2O_7$-1313 were grown by a floating zone method at University of Tennessee in a vertical optical-image furnace. Stoichiometric mixtures of $La_2O_3$ (pretreated at 1000 °C) and NiO were ground and fired at 1050 °C for 1 day. These precursor powders were hydrostatically pressed into a rod and sintered at 1400 °C for 12-24 hours. Crystals of $La_3Ni_2O_7$-2222 and $La_3Ni_2O_7$-1313 were grown directly from the sintered rods in 100% $O_2$ at a pressure of around 14-15 bars. During the crystal growth, the traveling rate was 3-4 mm/hour, and the feed rod and seed were counter-rotated at a rate in the range of 15-20 rpm. $La_3Ni_2O_7$-2222 and $La_3Ni_2O_7$-1313 crystals were obtained from some sections of the grown boules. The two kinds of crystals were identified by single crystal X-ray diffraction. No uncommon hazards are noted in this experimental procedure.

**Single crystal X-ray diffraction measurement**: The single crystal of $La_3Ni_2O_7$ was picked up, mounted on a nylon loop with paratone oil, and measured using a XtalLAB Synergy, Dualflex, Hypix single crystal X-ray diffractometer with an Oxford Cryosystems low-temperature device, operating at $T = 300(1)$ K and $T = 100(1)$ K. Data were measured using $\omega$ scans using Mo K$_\alpha$ radiation ($\lambda = 0.71073$ Å, micro-focus sealed X-ray tube, 50 kV, 1 mA). The total number of runs and images was based on the strategy calculation from the program CrysAlisPro 1.171.43.92a (Rigaku OD, 2023). Data reduction was performed with correction for Lorentz polarization. Numerical absorption correction based on gaussian integration over a multifaceted crystal model. Empirical absorption correction using spherical harmonics, implemented in SCALE3 ABSPACK scaling algorithm. The structure was solved and refined using the Bruker SHELXTL Software Package.[22,23]

**Electrical resistivity measurement**: Temperature-dependent electrical resistivity measurement was performed with a Quantum Design DynaCool physical property measurement system (PPMS) in the temperature range of 1.8–300 K at zero field with a four-probe method using platinum wires on a single crystal sample of $La_3Ni_2O_7$-1313 in the dimensions of 3.5 mm × 2.8 mm × 1.2 mm.

## Results and Discussion

**Crystal structure determination**: The crystal structures of Ruddlesden–Popper bilayer La$_3$Ni$_2$O$_{7-\delta}$ ($\delta$ = 0.00 and 0.08) were initially determined as $F$-centered orthorhombic $Fmmm$ through powder X-ray diffraction (XRD) and Rietveld refinement.[24] Recognizing the significant uncertainty in determining oxygen coordinates via XRD analysis, neutron powder diffraction (NPD) was subsequently employed.[25,26] The results obtained indicated that the $Fmmm$ space group was inappropriate, as it failed to account for extra weak peaks, hinting at a lower symmetry. This led to the proposal that the most suitable structure model involves a $C$-centered orthorhombic lattice of the space group $Cmcm$.[26] The challenges inherent in the pseudo $F$-centered orthorhombic lattice, the influence of oxygen vacancies on the structure, and the coexistence of Ruddlesden–Popper bilayer and trilayer phases underscore the complexities in structure determination, necessitating the growth of pure La$_3$Ni$_2$O$_7$ single crystals. In a recent development, the high-pressure floating zone method, enabling 100% O$_2$ atmosphere with controllable gas pressure, has successfully facilitated the single crystal growth of Ruddlesden–Popper nickelates with high purity.[27] This advancement opens possibilities for studying crystal structures using X-rays, which is more cost-effective and accessible than neutrons and provides precise enough structure determination results.

Here, utilizing this high-pressure floating zone method, high-purity single crystals of Ruddlesden–Popper bilayer nickelate La$_3$Ni$_2$O$_7$ (termed La$_3$Ni$_2$O$_7$-2222) have been obtained. Our single crystal XRD investigations confirm the presence of a pseudo $F$-centered orthorhombic lattice and the characteristic Ruddlesden–Popper bilayer stacking, as shown in **Figure 1**. The single crystal XRD refinement details are summarized in **Table 1** and **2**. No instances of oxygen vacancy have been observed with careful refinement attempts. The structural difference between the $F$-centered $Fmmm$ and $C$-centered $Cmcm$ models hinges on the positioning of oxygen atoms in the perovskite layers (labeled as O4 in La$_3$Ni$_2$O$_7$-2222 in **Figure 1** and **Figure 2a**). Specifically, this distinction lies in whether the oxygen resides on a symmetry-restricted special site ($Fmmm$) or a more general site ($Cmcm$). Of particular note is the characteristic observed in the out-of-plane Ni–O–Ni bond angle emphasized in **Figure 2**. In the $Cmmm$ configuration, this angle refines to be 168.3(3)°, a significant deviation from the 180° observed in the $Fmmm$ setting, providing additional evidence against the appropriateness of the *Fmmm* model. A recent report also suggests that this *C*-centered lattice becomes increasingly favorable as the temperature decreases.[28]

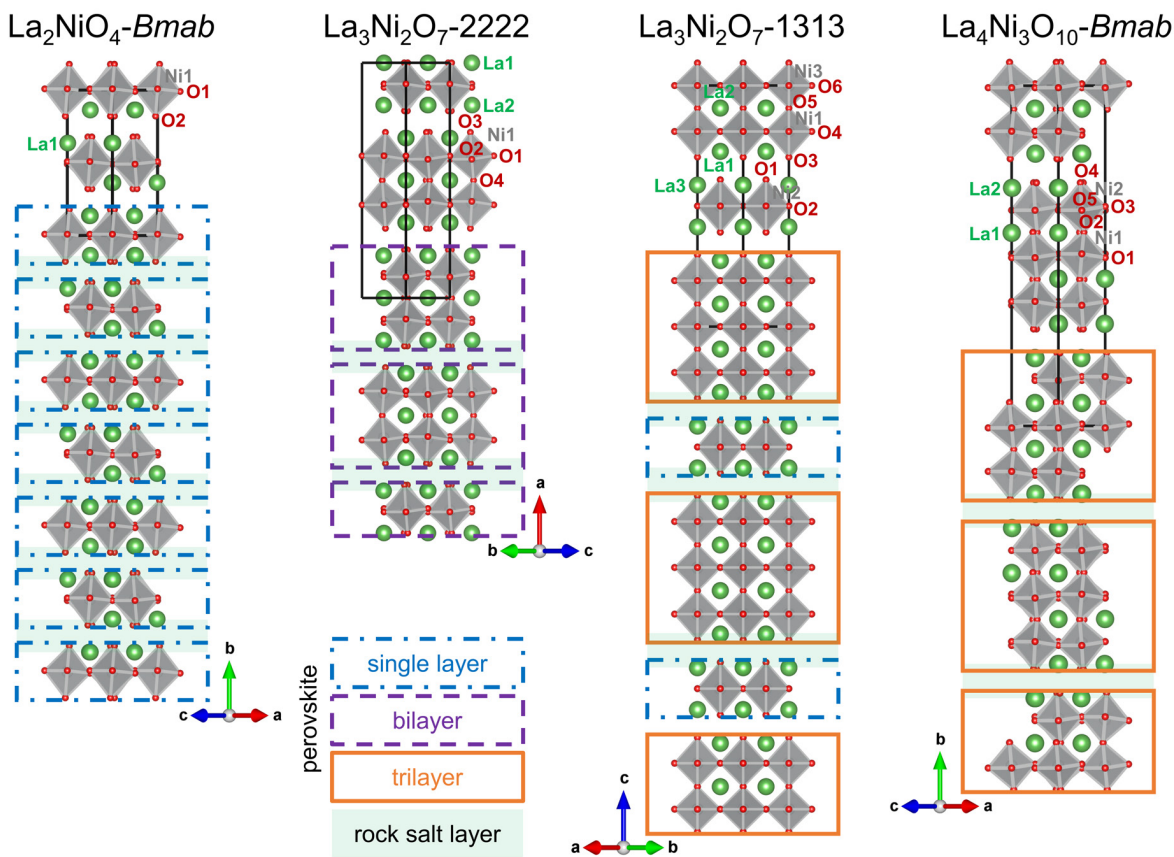

**Figure 1** View of layer stacking in Ruddlesden–Popper nickelates, single layer $La_2NiO_4$, bilayer $La_3Ni_2O_7$-2222, the hidden phase $La_3Ni_2O_7$-1313, and trilayer $La_4Ni_3O_{10}$. Green, grey, and red represent La, Ni, and O atoms. Crystallographically unique atoms are labeled. Perovskite layers are highlighted with blue dash-dot lines, purple dash lines, and orange lines, indicating $n$ = 1, 2, and 3, respectively. Rock salt layers are colored with light green.

Furthermore, our observations reveal the existence of a hidden phase in this system, termed $La_3Ni_2O_7$-1313, which adopts a strongly preferred *C*-centered orthorhombic lattice with space group *Cmmm*. In its structural arrangement, the perovskite [$LaNiO_3$] layers exhibit a systematic long-range order, alternating between single- and trilayer configurations (…1313…), as elucidated in **Figure 1**, represented by the identical chemical formula $La_3Ni_2O_7$ to bilayer $La_3Ni_2O_7$-2222. The single crystal XRD refinement details of $La_3Ni_2O_7$-1313 at 300 K and 100 K are summarized in **Table 1** and **3**, and **Table S1** and **S2**, respectively. Negligible vacancy has been observed regarding oxygen stoichiometry. **Figure 1** also presents the characteristic stacking of single-, bilayer, and trilayer structures in Ruddlesden–Popper nickelates.[26,29] A clear difference arises between $La_3Ni_2O_7$-1313 and $La_3Ni_2O_7$-2222 when considering perovskite layers and rock salt layers as the structure building blocks. The $La_3Ni_2O_7$-1313 structure is constructed through a long range order of single- and trilayer perovskite building blocks in $La_2NiO_4$ and $La_4Ni_3O_{10}$, respectively, without the bilayer counterparts present in $La_3Ni_2O_7$-2222. Recent research on the design and synthesis of a hybrid layered Ruddlesden–Popper nickelate by a flux method, featuring a …1212… sequence,[30] could provide a potential avenue for experimental validation.

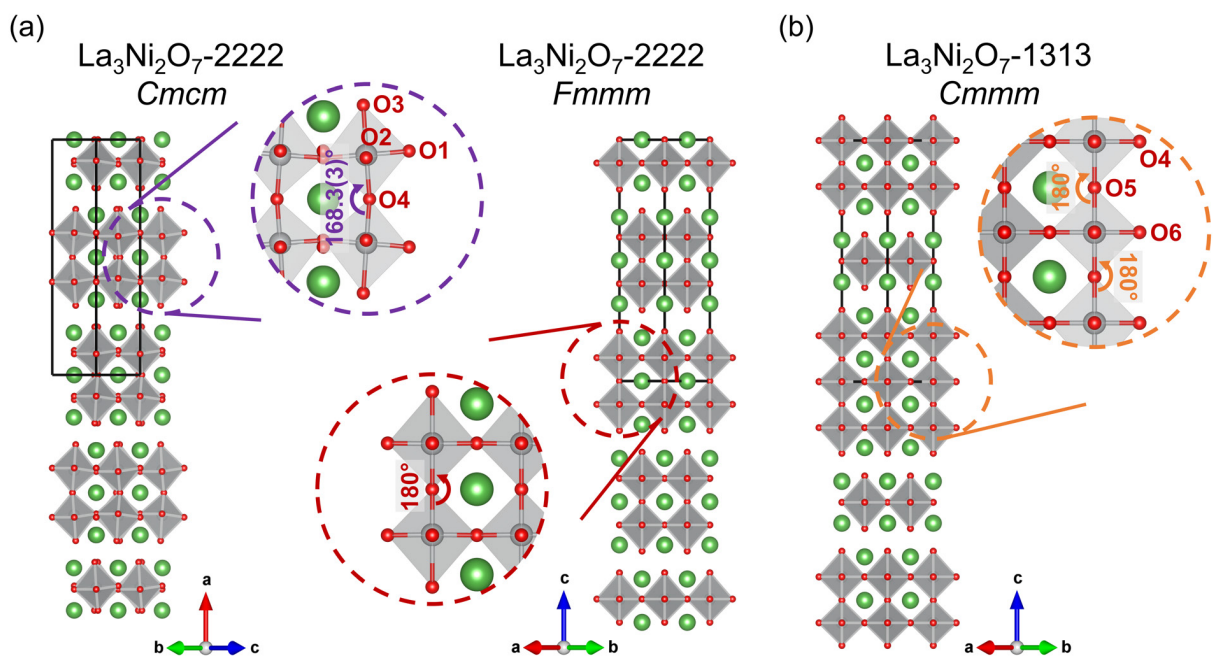

**Figure 2** View of the characteristic out-of-plane Ni–O–Ni bond angle in (**a**) bilayer $La_3Ni_2O_7$-2222 (*Cmcm* and *Fmmm* settings) and (**b**) the hidden phase $La_3Ni_2O_7$-1313. Green, grey, and red represent La, Ni, and O atoms. The corresponding oxygen atoms and bond angles are labeled as presented.

**Table 1** The crystal structure and refinement of $La_3Ni_2O_7$-2222 and $La_3Ni_2O_7$-1313 at 300 K.

| Chemical Formula | $La_3Ni_2O_7$-2222 | $La_3Ni_2O_7$-1313 |
|---|---|---|
| Formula weight | 646.15 g/mol | 646.15 g/mol |
| Space Group | $Cmcm$ | $Cmmm$ |
| Unit cell dimensions | $a$ = 20.5295(7) Å | $a$ = 5.4399(2) Å |
| | $b$ = 5.44607(17) Å | $b$ = 5.4594(2) Å |
| | $c$ = 5.3921(2) Å | $c$ = 20.3265(5) Å |
| Volume | 602.86(4) Å³ | 603.66(4) Å³ |
| $Z$ | 4 | 4 |
| Density (calculated) | 7.119 g/cm³ | 7.110 g/cm³ |
| Absorption coefficient | 26.916 mm⁻¹ | 26.881 mm⁻¹ |
| $F(000)$ | 1132 | 1132 |
| $2\theta$ range | 7.74 to 81.66° | 6.02 to 82.62° |
| Reflections collected | 18169 | 19216 |
| Independent reflections | 1059 [$R_{int}$ = 0.0627] | 1164 [$R_{int}$ = 0.0755] |
| Refinement method | Full-matrix least-squares on $F^2$ | Full-matrix least-squares on $F^2$ |
| Data / restraints / parameters | 1059 / 0 / 37 | 1164 / 0 / 49 |
| Final $R$ indices | $R_1$ (I>2σ(I)) = 0.0267; $wR_2$ (I>2σ(I)) = 0.0703 | $R_1$ (I>2σ(I)) = 0.0301; $wR_2$ (I>2σ(I)) = 0.0570 |
| | $R_1$ (all) = 0.0350; $wR_2$ (all) = 0.0742 | $R_1$ (all) = 0.0382; $wR_2$ (all) = 0.0585 |
| Largest diff. peak and hole | +6.124 e/Å⁻³ and −1.572 e/Å⁻³ | +2.316 e/Å⁻³ and −2.165 e/Å⁻³ |
| R.M.S. deviation from mean | 0.452 e/Å⁻³ | 0.353 e/Å⁻³ |
| Goodness-of-fit on $F^2$ | 1.130 | 1.304 |

**Table 2** Atomic coordinates and equivalent isotropic atomic displacement parameters (Å²) of $La_3Ni_2O_7$-2222 at 300 K. $U_{eq}$ is defined as one third of the trace of the orthogonalized $U_{ij}$ tensor.

| | Wyck. | $x$ | $y$ | $z$ | Occ. | $U_{eq}$ |
|---|---|---|---|---|---|---|
| **La₁** | 4$c$ | 0 | 0.75056(4) | 1/4 | 1 | 0.00746(8) |
| **La₂** | 8$g$ | 0.32019(2) | 0.25774(3) | 1/4 | 1 | 0.00636(7) |
| **Ni** | 8$g$ | 0.09588(3) | 0.25241(6) | 1/4 | 1 | 0.00470(10) |
| **O₁** | 8$e$ | 0.39563(15) | 0 | 0 | 1 | 0.0113(5) |
| **O₂** | 8$e$ | 0.08964(15) | 0 | 0 | 1 | 0.0111(5) |
| **O₃** | 8$g$ | 0.20433(15) | 0.21690(6) | 1/4 | 1 | 0.0131(5) |
| **O₄** | 4$c$ | 0 | 0.28950(8) | 1/4 | 1 | 0.0114(6) |

**Table 3** Atomic coordinates and equivalent isotropic atomic displacement parameters ($\text{Å}^2$) of $La_3Ni_2O_7$-1313 at 300 K. $U_{eq}$ is defined as one third of the trace of the orthogonalized $U_{ij}$ tensor.

|  | Wyck. | $x$ | $y$ | $z$ | Occ. | $U_{eq}$ |
|---|---|---|---|---|---|---|
| La$_1$ | 4$l$ | 0 | 1/2 | 0.27343(2) | 1 | 0.00909(8) |
| La$_2$ | 4$l$ | 0 | 1/2 | 0.09348(2) | 1 | 0.00953(8) |
| La$_3$ | 4$k$ | 0 | 0 | 0.41317(2) | 1 | 0.00914(8) |
| Ni$_1$ | 4$k$ | 0 | 0 | 0.19063(4) | 1 | 0.00627(13) |
| Ni$_2$ | 2$c$ | 1/2 | 0 | 1/2 | 1 | 0.00779(19) |
| Ni$_3$ | 2$a$ | 0 | 0 | 0 | 1 | 0.00550(18) |
| O$_1$ | 4$l$ | 0 | 1/2 | 0.39120(3) | 1 | 0.0300(17) |
| O$_2$ | 4$f$ | 1/4 | 1/4 | 1/2 | 1 | 0.0132(9) |
| O$_3$ | 4$k$ | 0 | 0 | 0.29700(3) | 1 | 0.0277(16) |
| O$_4$ | 8$m$ | 1/4 | 1/4 | 0.19070(2) | 1 | 0.0179(8) |
| O$_5$ | 4$k$ | 0 | 0 | 0.09380(3) | 1 | 0.0341(18) |
| O$_6$ | 4$e$ | 1/4 | 1/4 | 0 | 1 | 0.039(2) |

**Table 4** provides a summary of Ni–O bond lengths and Ni–O–Ni bond angles in the structure of $La_3Ni_2O_7$-1313. A noteworthy observation is the larger distortion of [$NiO_6$] octahedra in the outer Ni of the trilayer (Ni1 in **Figure 1**) in comparison to the inner Ni of the trilayer (Ni3) and Ni in the single layer (Ni2). This aligns with reports in other multilayer Ruddlesden–Popper phases.[31] Meanwhile, the Ni–O bond lengths within the perovskite layers on the outer side (Ni1–O3 and Ni2–O1) are approximately 10% greater than Ni–O inside (Ni1–O5 and Ni3–O5) in $La_3Ni_2O_7$-1313. To evaluate the valence states of three crystallographically unique Ni in the structure, bond valence sums were calculated ($R_0 = 1.689$, $B = 0.347$)[32]. The obtained values for Ni in the single layer, and inner and outer Ni in the trilayer reveal differences, indicating charge differentiation among them and providing evidence for potential charge transfer in the system. Remarkably, the bond valence sums of the inner and outer Ni in the trilayer of $La_3Ni_2O_7$-1313 exhibit no significant difference from those reported in $La_4Ni_3O_{10}$,[27] suggesting a connection between trilayer building blocks.

Our experimental reciprocal lattice planes are presented in **Figure 3**. To enable a clearer visual comparison between $La_3Ni_2O_7$-1313 and $La_3Ni_2O_7$-2222, the *Cmcm* unit cell (details in **Table 1**) of $La_3Ni_2O_7$-2222 was transformed to a symmetry-equivalent *Amam* setting (*c* axis perpendicular to perovskite layers). Begin with $La_3Ni_2O_7$-1313, given its space group *Cmmm*, reflections at $h + k = 2n$ would be expected due to *C*-centering, a condition met by all involved reflections. In contrast, for $La_3Ni_2O_7$-2222, besides *A*-centering ($k + l = 2n$), the presence of the *a* glide plane perpendicular to *b* axis results in absences in the (*h*0*l*) reciprocal plane when $h = 2n + 1$. Thus, the overall reflection conditions in the (*h*0*l*) plane are defined by $h = 2n$ and $l = 2n$. Consequently, when examining the (*h*0*l*) plane, we expect to see twice as many reflections along *c** when $h = 2n$ in $La_3Ni_2O_7$-1313 compared to $La_3Ni_2O_7$-2222. **Figure 3a** and **3c** provide a visual representation of the distinct reflection conditions due to lattice centering, while **Figure 3b** and **3d** precisely illustrate the two distinct space groups in $La_3Ni_2O_7$-2222 and $La_3Ni_2O_7$-1313 as discussed above. Additionally, the (0*kl*) reciprocal planes in **Figure S1** further validate the differences in reflection conditions because of lattice centering.

However, we also noted weak reflections at half-integer positions within current reciprocal lattice in the (0*kl*) plane in $La_3Ni_2O_7$-1313, shown in **Figure S2**. This observation may suggest a lower symmetry or a mixture of *Cmmm* and the lower symmetry, as mentioned in a recent report on $La_3Ni_2O_7$-1313.[33] It is crucial to note that this observation does not compromise the distinctive …1313… stacking sequence. The precise determination has been somewhat restricted by home lab single crystal XRD and may necessitate further examination through synchrotron X-ray and neutron diffraction.

**Table 4** Selected bond lengths and bond angles in $La_3Ni_2O_7$-1313. Crystallographically unique atoms are labeled in **Figure 1**.

|  | $La_3Ni_2O_7$-1313 |
| --- | --- |
| Ni–O (single layer) / Å | Ni2–O1: 2.211(6) |
|  | Ni2–O2: 1.92674(5) |
| Bond valence sum | 2.46 |
| Ni–O (outer Ni of the trilayer) / Å | Ni1–O3: 2.162(6) |
|  | Ni1–O4: 1.92674(5) |
|  | Ni1–O5: 1.969(6) |
| Bond valence sums | 2.72 |
| Ni–O (inner Ni of the trilayer) / Å | Ni3–O5: 1.906(6) |
|  | Ni3–O6: 1.92674(5) |
| Bond valence sum | 3.09 |
| In-plane Ni–O–Ni / deg | Ni2–O2–Ni2: 180 |
|  | Ni1–O4–Ni1: 179.9(3) |
|  | Ni3–O6–Ni3: 180 |
| Out-of-plane Ni–O–Ni /deg | Ni1–O5–Ni3: 180 |

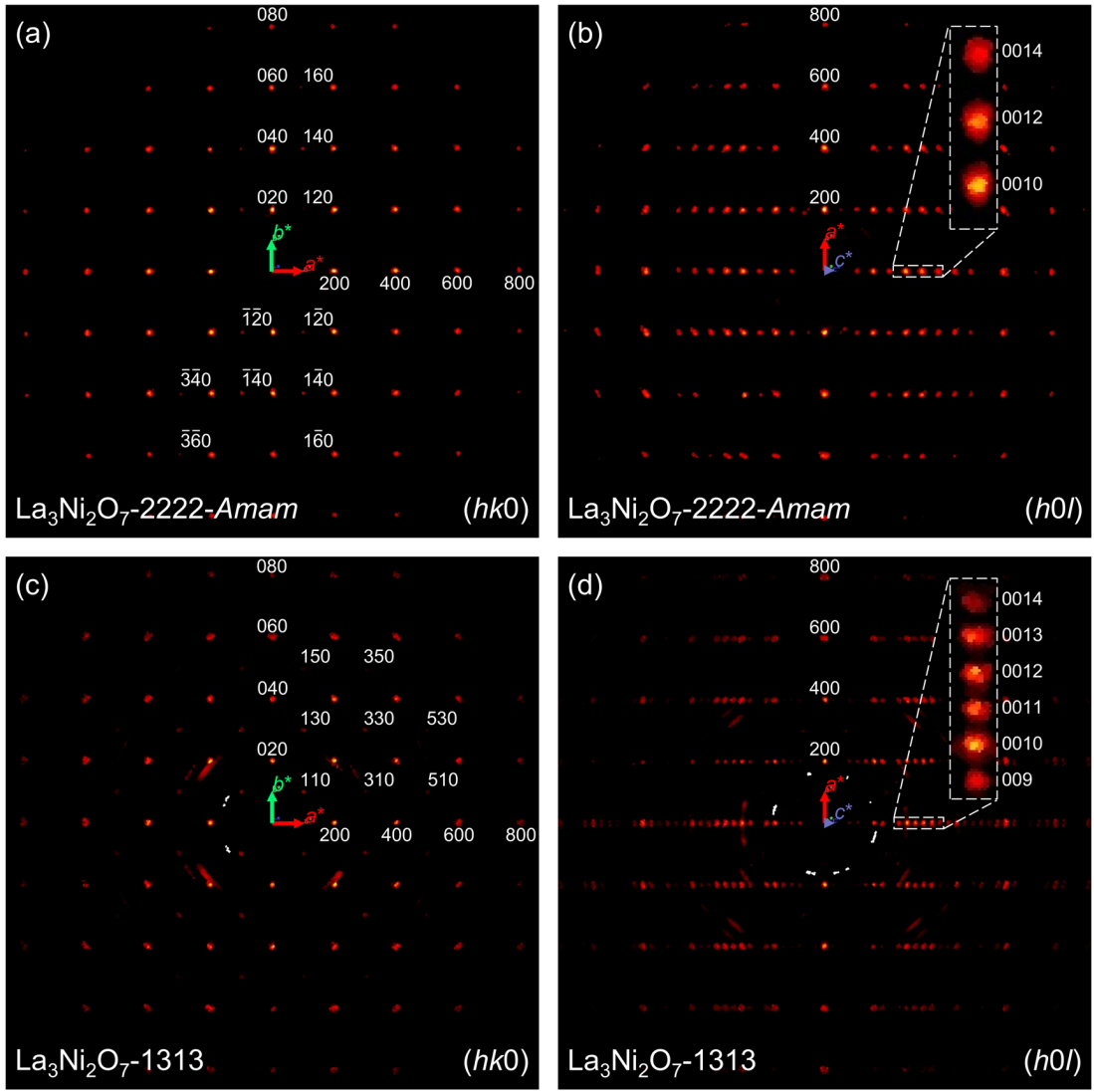

**Figure 3** Reciprocal lattice planes of La$_3$Ni$_2$O$_7$-1313 and La$_3$Ni$_2$O$_7$-2222. (***a, b***) ($hk$0) and ($h0l$) planes of La$_3$Ni$_2$O$_7$-2222. The *Cmcm* unit cell was transformed to *Amam* setting for a better visual comparison with La$_3$Ni$_2$O$_7$-1313. (***c, d***) ($hk$0) and ($h0l$) planes of La$_3$Ni$_2$O$_7$-1313.

**Electrical resistivity measurement**: The temperature dependence of electrical resistivity and its temperature derivative of $La_3Ni_2O_7$-1313 in the range of 2–300 K was presented in **Figure 4**. The resistivity at 300 K measures approximately 0.035 $\Omega$ cm, a value small enough to preclude the presence of significant contact resistance. Our measurements indicate that the sample exhibit a semiconductor-like behavior, a departure from the characteristics reported for $La_3Ni_2O_7$-2222.[17] A kink at about 50 K was observed in our resistivity curve, which was attributed to unstable contacts. It is essential to emphasize that this kink should not be accounted for the intrinsic properties of $La_3Ni_2O_7$-1313 but rather considered a result of contact instability. Two very recent reports on $La_3Ni_2O_7$-1313 notes metallic behavior at ambient pressure,[33,34] suggesting sample-dependent electrical resistivity. Considering the structure building blocks derived from Ruddlesden–Popper single layer $La_2NiO_4$ and trilayer $La_4Ni_3O_{10}$ in $La_3Ni_2O_7$-1313, we propose that the efficiency of charge-transfer between Ni in these single- and trilayer perovskite layers may dictate the bulk electrical resistivity behaviors of this system.

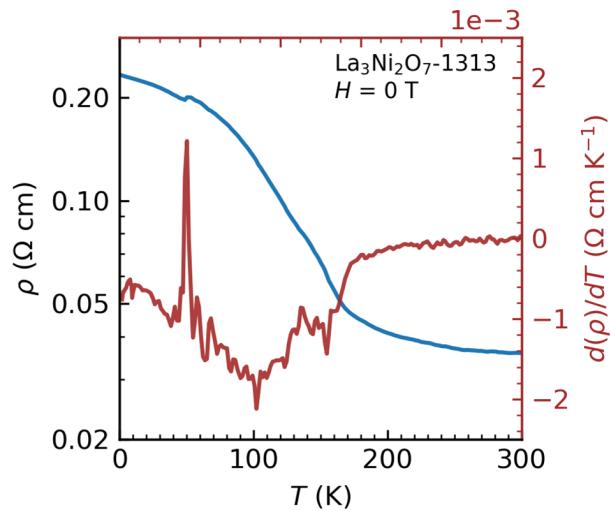

**Figure 4** Temperature-dependent electrical resistivity and its temperature derivative of $La_3Ni_2O_7$-1313. The kink at ~50 K comes from unstable contacts.

## Conclusion

In conclusion, we present the crystal structure and electrical resistivity of two Ruddlesden-Popper nickelates: bilayer $La_3Ni_2O_7$-2222, and the hidden phase, $La_3Ni_2O_7$-1313. Our single crystal XRD measurements confirm that $La_3Ni_2O_7$-2222 features the pseudo *F*-centered orthorhombic lattice with bilayer perovskite [$LaNiO_3$] layers separated by rock salt [$LaO$] layers (…2222…). Remarkably, the semiconductor-like $La_3Ni_2O_7$-1313 adopts a strongly preferred *C*-centered orthorhombic lattice with the space group *Cmmm*. In the structure, perovskite [$LaNiO_3$] layers exhibit a systematic long-range order, alternating between single- and trilayer configurations (…1313…). This report introduces new possibilities for exploring the crystal structure and the structure-property relationship in Ruddlesden–Popper nickelates.

## Associated Content

Supporting Information Available: Reciprocal lattice planes, (0*kl*), of $La_3Ni_2O_7$-2222 and $La_3Ni_2O_7$-1313; Zoom-in (0*kl*) plane of $La_3Ni_2O_7$-1313; The crystal structure and refinement of $La_3Ni_2O_7$-1313 at 100 K; Atomic coordinates and equivalent isotropic atomic displacement parameters of $La_3Ni_2O_7$-1313 at 100 K.

This information is available free of charge at the website: http://pubs.acs.org/

## Acknowledgments

The work at Michigan State University was supported by U.S. DOE-BES under Contract DE-SC0023648. The work at University of Tennessee (crystal growth and electrical resistivity measurement) was supported by the Air Force Office of Scientific Research under Grant No. FA9550-23-1-0502. H.W. and W.X. appreciate helpful discussions with Dr. Xinglong Chen and Dr. J.F. Mitchell (Argonne National Laboratory).

# References

1.  Bednorz, J. G.; Müller, K. A., Possible highTc superconductivity in the Ba−La−Cu−O system. *Z. Phys. B: Condens. Matter* **1986**, *64* (2), 189-193.

2.  Anisimov, V. I.; Bukhvalov, D.; Rice, T. M., Electronic structure of possible nickelate analogs to the cuprates. *Phys. Rev. B* **1999**, *59* (12), 7901-7906.

3.  Ruddlesden, S. N.; Popper, P., New compounds of the $K_2NiF_4$ type. *Acta Crystallogr.* **1957**, *10* (8), 538-539.

4.  Ruddlesden, S. N.; Popper, P., The compound $Sr_3Ti_2O_7$ and its structure. *Acta Crystallogr.* **1958**, *11* (1), 54-55.

5.  Crespin, M.; Levitz, P.; Gatineau, L., Reduced forms of $LaNiO_3$ perovskite. Part 1.— Evidence for new phases: $La_2Ni_2O_5$ and $LaNiO_2$. *J. Chem. Soc., Faraday Trans.* **1983**, *79* (8), 1181-1194.

6.  Lacorre, P., Passage from T-type to T'-type arrangement by reducing $R_4Ni_3O_{10}$ to $R_4Ni_3O_8$ ($R$ = La, Pr, Nd). *J. Solid State Chem.* **1992**, *97* (2), 495-500.

7.  Hayward, M. A.; Rosseinsky, M. J., Synthesis of the infinite layer Ni(I) phase $NdNiO_{2+x}$ by low temperature reduction of $NdNiO_3$ with sodium hydride. *Solid State Sci.* **2003**, *5* (6), 839-850.

8.  Crespin, M.; Isnard, O.; Dubois, F.; Choisnet, J.; Odier, P., $LaNiO_2$: Synthesis and structural characterization. *J. Solid State Chem.* **2005**, *178* (4), 1326-1334.

9.  Poltavets, V. V.; Lokshin, K. A.; Dikmen, S.; Croft, M.; Egami, T.; Greenblatt, M., $La_3Ni_2O_6$: A New Double T'-type Nickelate with Infinite $Ni^{1+/2+}O_2$ Layers. *J. Am. Chem. Soc.* **2006**, *128* (28), 9050-9051.

10. Li, D.; Lee, K.; Wang, B. Y.; Osada, M.; Crossley, S.; Lee, H. R.; Cui, Y.; Hikita, Y.; Hwang, H. Y., Superconductivity in an infinite-layer nickelate. *Nature* **2019**, *572* (7771), 624-627.

11. Osada, M.; Wang, B. Y.; Goodge, B. H.; Lee, K.; Yoon, H.; Sakuma, K.; Li, D.; Miura, M.; Kourkoutis, L. F.; Hwang, H. Y., A Superconducting Praseodymium Nickelate with Infinite Layer Structure. *Nano Lett.* **2020**, *20* (8), 5735-5740.

12. Wang, N. N.; Yang, M. W.; Yang, Z.; Chen, K. Y.; Zhang, H.; Zhang, Q. H.; Zhu, Z. H.; Uwatoko, Y.; Gu, L.; Dong, X. L.; Sun, J. P.; Jin, K. J.; Cheng, J. G., Pressure-induced monotonic enhancement of $T_c$ to over 30 K in superconducting $Pr_{0.82}Sr_{0.18}NiO_2$ thin films. *Nat. Commun.* **2022**, *13* (1), 4367.

13. Osada, M.; Wang, B. Y.; Goodge, B. H.; Harvey, S. P.; Lee, K.; Li, D.; Kourkoutis, L. F.; Hwang, H. Y., Nickelate Superconductivity without Rare-Earth Magnetism: (La,Sr)NiO$_2$. *Adv. Mater.* **2021**, *33* (45), 2104083.

14. Zeng, S.; Li, C.; Chow, L. E.; Cao, Y.; Zhang, Z.; Tang, C. S.; Yin, X.; Lim, Z. S.; Hu, J.; Yang, P.; Ariando, A., Superconductivity in infinite-layer nickelate $La_{1-x}Ca_xNiO_2$ thin films. *Sci. Adv.* **2022**, *8* (7), eabl9927.

15. Pan, G. A.; Ferenc Segedin, D.; LaBollita, H.; Song, Q.; Nica, E. M.; Goodge, B. H.; Pierce, A. T.; Doyle, S.; Novakov, S.; Córdova Carrizales, D.; N'Diaye, A. T.; Shafer, P.; Paik, H.; Heron, J. T.; Mason, J. A.; Yacoby, A.; Kourkoutis, L. F.; Erten, O.; Brooks, C. M.; Botana, A. S.; Mundy, J. A., Superconductivity in a quintuple-layer square-planar nickelate. *Nat. Mater.* **2022**, *21* (2), 160-164.

16. Ferenc Segedin, D.; Goodge, B. H.; Pan, G. A.; Song, Q.; LaBollita, H.; Jung, M.-C.; El-Sherif, H.; Doyle, S.; Turkiewicz, A.; Taylor, N. K.; Mason, J. A.; N'Diaye, A. T.; Paik, H.; El Baggari, I.; Botana, A. S.; Kourkoutis, L. F.; Brooks, C. M.; Mundy, J. A., Limits to the strain engineering of layered square-planar nickelate thin films. *Nat. Commun.* **2023**, *14* (1), 1468.

17. Sun, H.; Huo, M.; Hu, X.; Li, J.; Liu, Z.; Han, Y.; Tang, L.; Mao, Z.; Yang, P.; Wang, B.; Cheng, J.; Yao, D.-X.; Zhang, G.-M.; Wang, M., Signatures of superconductivity near 80 K in a nickelate under high pressure. *Nature* **2023**, *621* (7979), 493-498.

18. Hou, J.; Yang, P.-T.; Liu, J.-Y.; Shan, P.-F.; Ma, L.; Wang, G.; Wang, N.-N.; Guo, H.-Z.; Sun, J.-P.; Uwatoko, Y.; Wang, M.; Zhang, G.-M.; Wang, B.-S.; Cheng, J.-G., Emergence of High-Temperature Superconducting Phase in Pressurized $La_3Ni_2O_7$ Crystals. *Chin. Phys. Lett.* **2023**, *40* (11), 117302.

19. Zhang, Y.; Su, D.; Huang, Y.; Sun, H.; Huo, M.; Shan, Z.; Ye, K.; Yang, Z.; Li, R.; Smidman, M.; Wang, M.; Jiao, L.; Yuan, H., High-temperature superconductivity with zero-resistance and strange metal behavior in $La_3Ni_2O_7$. **2023**, arXiv:2307.14819.

20. Wang, G.; Wang, N.; Hou, J.; Ma, L.; Shi, L.; Ren, Z.; Gu, Y.; Shen, X.; Ma, H.; Yang, P.; Liu, Z.; Guo, H.; Sun, J.; Zhang, G.; Yan, J.; Wang, B.; Uwatoko, Y.; Cheng, J., Pressure-induced superconductivity in polycrystalline $La_3Ni_2O_7$. **2023**, arXiv:2309.17378.

21. Zhang, M.; Pei, C.; Wang, Q.; Zhao, Y.; Li, C.; Cao, W.; Zhu, S.; Wu, J.; Qi, Y., Effects of Pressure and Doping on Ruddlesden-Popper phases La$_{n+1}$Ni$_n$O$_{3n+1}$. **2023**, arXiv:2309.01651.

22. Sheldrick, G., SHELXT - Integrated space-group and crystal-structure determination. *Acta Cryst. A* **2015**, *71* (1), 3-8.

23. Sheldrick, G., Crystal structure refinement with SHELXL. *Acta Cryst. C* **2015**, *71* (1), 3-8.

24. Zhang, Z.; Greenblatt, M.; Goodenough, J. B., Synthesis, Structure, and Properties of the Layered Perovskite La$_3$Ni$_2$O$_{7-\delta}$. *J. Solid State Chem.* **1994**, *108* (2), 402-409.

25. Voronin, V. I.; Berger, I. F.; Cherepanov, V. A.; Gavrilova, L. Y.; Petrov, A. N.; Ancharov, A. I.; Tolochko, B. P.; Nikitenko, S. G., Neutron diffraction, synchrotron radiation and EXAFS spectroscopy study of crystal structure peculiarities of the lanthanum nickelates La$_{n+1}$Ni$_n$O$_y$ ($n$ = 1, 2, 3). *Nucl. Instrum. Methods Phys. Res. A.* **2001**, *470* (1), 202-209.

26. Ling, C. D.; Argyriou, D. N.; Wu, G.; Neumeier, J. J., Neutron Diffraction Study of La$_3$Ni$_2$O$_7$: Structural Relationships Among $n$ = 1, 2, and 3 Phases La$_{n+1}$Ni$_n$O$_{3n+1}$. *J. Solid State Chem.* **2000**, *152* (2), 517-525.

27. Zhang, J.; Zheng, H.; Chen, Y.-S.; Ren, Y.; Yonemura, M.; Huq, A.; Mitchell, J. F., High oxygen pressure floating zone growth and crystal structure of the metallic nickelates $R_4$Ni$_3$O$_{10}$ ($R$ = La, Pr). *Phys. Rev. Mater.* **2020**, *4* (8), 083402.

28. Xu, M.; Huyan, S.; Wang, H.; Bud'ko, S. L.; Chen, X.; Ke, X.; Mitchell, J. F.; Canfield, P. C.; Li, J.; Xie, W., Pressure-dependent "Insulator-Metal-Insulator" Behavior in Sr-doped La$_3$Ni$_2$O$_7$. **2023**, arXiv:2312.14251.

29. Medarde, M.; Rodríguez-Carvajal, J., Oxygen vacancy ordering in La$_{2-x}$Sr$_x$NiO$_{4-\delta}$ ($0 \leqslant x \leqslant 0.5$): the crystal structure and defects investigated by neutron diffraction. *Z. Phys. B: Condens. Matter* **1997**, *102* (3), 307-315.

30. Li, F.; Guo, N.; Zheng, Q.; Shen, Y.; Wang, S.; Cui, Q.; Liu, C.; Wang, S.; Tao, X.; Zhang, G.-M.; Zhang, J., Design and synthesis of three-dimensional hybrid Ruddlesden-Popper nickelate single crystals. **2023**, arXiv:2312.08116.

31. Mitchell, J. F.; Argyriou, D. N.; Jorgensen, J. D.; Hinks, D. G.; Potter, C. D.; Bader, S. D., Charge delocalization and structural response in layered La$_{1.2}$Sr$_{1.8}$Mn$_2$O$_7$: Enhanced distortion in the metallic regime. *Phys. Rev. B* **1997**, *55* (1), 63-66.

32. Gagne, O. C.; Hawthorne, F. C., Comprehensive derivation of bond-valence parameters for ion pairs involving oxygen. *Acta Cryst. B* **2015**, *71* (5), 562-578.

33. Chen, X.; Zhang, J.; Thind, A. S.; Sharma, S.; LaBollita, H.; Peterson, G.; Zheng, H.; Phelan, D. P.; Botana, A. S.; Klie, R. F.; Mitchell, J. F., Polymorphism in the Ruddlesden–Popper Nickelate La3Ni2O7: Discovery of a Hidden Phase with Distinctive Layer Stacking. *J. Am. Chem. Soc.* **2024**.

34. Puphal, P.; Reiss, P.; Enderlein, N.; Wu, Y.-M.; Khaliullin, G.; Sundaramurthy, V.; Priessnitz, T.; Knauft, M.; Richter, L.; Isobe, M.; van Aken, P. A.; Takagi, H.; Keimer, B.; Eren Suyolcu, Y.; Wehinger, B.; Hansmann, P.; Hepting, M., Unconventional crystal structure of the high-pressure superconductor $La_3Ni_2O_7$. **2023**, arXiv:2312.07341.

**For Table of Contents Only**

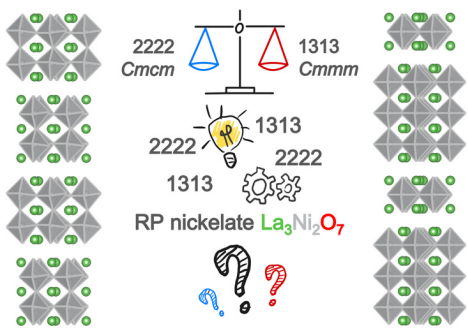

This study delves into the crystallographic and electrical properties of two polymorphs of Ruddlesden-Popper nickelates La$_3$Ni$_2$O$_7$: bilayer 2222-phase, and a hidden 1313-phase. The 1313-phase demonstrates a *C*-centered orthorhombic lattice (*Cmmm*, #65) with long-range ordering alternating between single layer and trilayer (…1313… sequence). This unique layer stacking shed light on the structure-property relationship in La$_3$Ni$_2$O$_7$, opening avenues for further exploration of their physical properties.

**Supporting Information**

# Long-Range Structural Order in a Hidden Phase of Ruddlesden–Popper Bilayer Nickelate La$_3$Ni$_2$O$_7$

Haozhe Wang[1], Long Chen[2], Aya Rutherford[2], Haidong Zhou[2], Weiwei Xie[1]*

1. Department of Chemistry, Michigan State University, East Lansing, MI, 48824, USA
2. Department of Physics and Astronomy, University of Tennessee, Knoxville, TN, 37996, USA

* Email: xieweiwe@msu.edu

## Table of Contents

**Figure S1** Reciprocal lattice planes, (0*kl*), of (***a***) La$_3$Ni$_2$O$_7$-2222 and (***b***) La$_3$Ni$_2$O$_7$-1313. At $k = 2n$, we observe twice as many reflections along $c^*$.

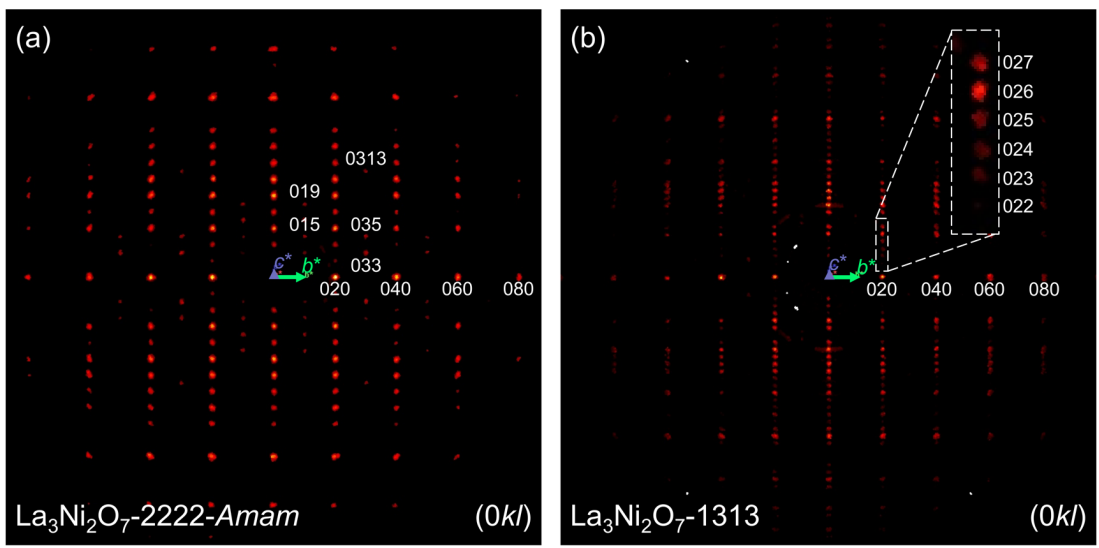

**Figure S2** Zoom-in (0*kl*) plane of La$_3$Ni$_2$O$_7$-1313. Weak reflections at half-integer positions have been observed.

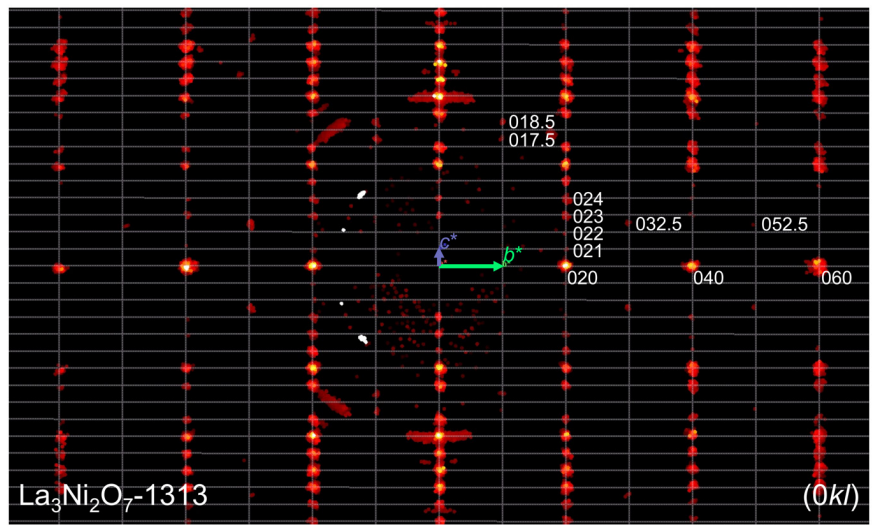

**Table S1** The crystal structure and refinement of $La_3Ni_2O_7$-1313 at 100 K.

| Chemical Formula | $La_3Ni_2O_7$-1313 |
|---|---|
| Formula weight | 646.15 g/mol |
| Space Group | $Cmmm$ |
| | $a$ = 5.4265(2) Å |
| Unit cell dimensions | $b$ = 5.4521(2) Å |
| | $c$ = 20.2969(6) Å |
| Volume | 600.50(4) Å$^3$ |
| $Z$ | 4 |
| Density (calculated) | 7.147 g/cm$^3$ |
| Absorption coefficient | 27.022 mm$^{-1}$ |
| $F(000)$ | 1132 |
| $2\theta$ range | 6.02 to 81.48° |
| Reflections collected | 14765 |
| Independent reflections | 1123 [$R_{int}$ = 0.0568] |
| Refinement method | Full-matrix least-squares on $F^2$ |
| Data / restraints / parameters | 1123 / 0 / 48 |
| Final $R$ indices | $R_1$ (I>2σ(I)) = 0.0493; $wR_2$ (I>2σ(I)) = 0.0967 |
| | $R_1$ (all) = 0.0547; $wR_2$ (all) = 0.0982 |
| Largest diff. peak and hole | +4.223 e/Å$^{-3}$ and –8.600 e/Å$^{-3}$ |
| R.M.S. deviation from mean | 0.543 e/Å$^{-3}$ |
| Goodness-of-fit on F$^2$ | 1.439 |

**Table S2** Atomic coordinates and equivalent isotropic atomic displacement parameters (Å$^2$) of $La_3Ni_2O_7$-1313 at 100 K. $U_{eq}$ is defined as one third of the trace of the orthogonalized $U_{ij}$ tensor.

| | Wyck. | $x$ | $y$ | $z$ | Occ. | $U_{eq}$ |
|---|---|---|---|---|---|---|
| **La$_1$** | 4$k$ | 0 | 0 | 0.22638(4) | 1 | 0.00704(13) |
| **La$_2$** | 4$l$ | 0 | 1/2 | 0.08680(3) | 1 | 0.00684(12) |
| **La$_3$** | 4$k$ | 0 | 0 | 0.40619(4) | 1 | 0.00837(13) |
| **Ni$_1$** | 4$l$ | 0 | 1/2 | 0.30907(8) | 1 | 0.0054(3) |
| **Ni$_2$** | 2$a$ | 0 | 0 | 0 | 1 | 0.0064(4) |
| **Ni$_3$** | 2$c$ | 0 | 1/2 | 1/2 | 1 | 0.0048(3) |
| **O$_1$** | 4$e$ | 1/4 | 1/4 | 0 | 1 | 0.0113(17) |
| **O$_2$** | 8$m$ | 1/4 | 1/4 | 0.3093(4) | 1 | 0.0156(14) |
| **O$_3$** | 4$l$ | 0 | 1/2 | 0.2021(6) | 1 | 0.019(2) |
| **O$_4$** | 4$f$ | 1/4 | 1/4 | 1/2 | 1 | 0.046(5) |
| **O$_5$** | 4$l$ | 0 | 1/2 | 0.4058(6) | 1 | 0.036(4) |
| **O$_6$** | 4$k$ | 0 | 0 | 0.1082(6) | 1 | 0.029(3) |